\documentclass[aps,showpacs,twocolumn,superscriptaddress]{revtex4}
\usepackage[dvips]{graphicx}
\usepackage{amsmath}
\usepackage{color}

\begin{document}

\title{Temperature dependence of thermal conductivities of coupled rotator lattice and the momentum diffusion in standard map}

\author{Yunyun Li}
\affiliation{Center for Phononics and
Thermal Energy Science and School of Physics Science and
Engineering, Tongji University, 200092 Shanghai, People's Republic
of China}

\author{Nianbei Li}
\email{nbli@tongji.edu.cn} \affiliation{Center for Phononics and
Thermal Energy Science and School of Physics Science and
Engineering, Tongji University, 200092 Shanghai, People's Republic
of China}

\author{Baowen Li}
\email{phononics@tongji.edu.cn} \affiliation{Center for Phononics
and Thermal Energy Science and School of Physics Science and
Engineering, Tongji University, 200092 Shanghai, People's Republic
of China} \affiliation{Department of Physics and Centre for
Computational Science and Engineering, National University of
Singapore, Singapore 117546, Republic of Singapore} \affiliation{NUS
Graduate School for Integrative Sciences and Engineering, Singapore
117456, Republic of Singapore}

\begin{abstract}
In contrary to other 1D momentum-conserving lattices such as the Fermi-Pasta-Ulam $\beta$ (FPU-$\beta$) lattice, the 1D coupled rotator lattice is a notable exception which conserves total momentum while exhibits normal heat conduction behavior. The temperature behavior of the thermal conductivities of 1D coupled rotator lattice had been studied in previous works trying to reveal the underlying physical mechanism for normal heat conduction. However, two different temperature behaviors of thermal conductivities have been claimed for the same coupled rotator lattice. These different temperature behaviors also intrigue the debate whether there is a phase transition of thermal conductivities as the function of temperature. In this work, we will revisit the temperature dependent thermal conductivities for the 1D coupled rotator lattice. We find that the temperature dependence follows a power law behavior which is different with the previously found temperature behaviors. Our results also support the claim that there is no phase transition for 1D coupled rotator lattice. We also give some discussion about the similarity of diffusion behaviors between the 1D coupled rotator lattice and the single kicked rotator also called the Chirikov standard map.
\end{abstract}

\maketitle
\section{Introduction}
The exploration of underlying mechanism for anomalous and normal heat conduction in low dimensional systems represents a huge challenge in the area of statistical physics~\cite{Lepri1997prl,Lepri2003pr,Dhar2008ap,Liu2013epjb}. After enormous efforts for more than one decade from numerical simulations\cite{Hu1998pre,Tong1999prb,Hatano1999pre,Tsironis1999pre,Sarmiento1999pre,Dhar1999prl,Alonso1999prl,Hu2000pre,Aoki2000pla,Li2001prl,Dhar2001prl,Aoki2001prl,Zhang2002pre,
Li2002prl,Saito2003epl,Savin2003pre,Lepri2003pre,Segal2003jcp,Gendelman2004prl,Li2005chaos,Zhang2005jcp,Zhao2005prl,
Zhao2006prl,He2008pre,Shao2008pre,Dubi2009pre,Henry2009prb,Saito2010prl,Wang2010prl,Yang2010nt,Wang2011epl,Wang2012pre,Xiong2012pre,
Landi2013pre,Xiong2014pre,Das2014pre,Mendl2014pre,Savin2014pre,Liu2014prb,Wang2015pre}, theoretical predictions\cite{Lepri1998epl,
Lepri1998pre,Narayan2002prl,Wang2004prl,Cipriani2005prl,Pereira2006prl,Delfini2006pre,Basile2006prl,Beijeren2012prl,
Mendl2013prl,Pereira2013pre,Spohn2014jsp,Liu2014prl} and experimental observations\cite{Chang2008prl,Xu2014nc,Meier2014prl}, there is still no consensus for the exact physical mechanism causing anomalous heat conduction. It is believed that momentum conservation plays an important role in determining the actual heat conduction behavior. In general, 1D non-integrable lattices with momentum conserving property should have anomalous heat conduction where the thermal conductivity $\kappa$ diverges with the lattice size $N$ as $\kappa\propto N^{\alpha}$ where $0<\alpha<1$~\cite{Lepri1997prl,Lepri2003pr,Dhar2008ap,Liu2013epjb}. However, the 1D coupled rotator lattice is a well known exception. It exhibits normal heat conduction behavior despite its momentum conserving nature~\cite{Giardina2000prl,Gendelman2000prl}.

The normal heat conduction in 1D coupled rotator lattice was discovered via numerical simulations by two groups independently~\cite{Giardina2000prl,Gendelman2000prl}. In order to understand the underlying mechanism, the temperature dependence of thermal conductivity has been studied in detail in both works. In Ref. \cite{Giardina2000prl}, the temperature dependence of thermal conductivity of coupled rotator lattice has been found to be like $\kappa(T)\propto e^{\Delta V/T}$ where $\Delta V$ is a positive constant. However, in Ref. \cite{Gendelman2000prl}, a temperature dependence of thermal conductivity of $\kappa(T)\propto e^{-T/A}$ was claimed where $A$ is a positive constant. There it was also argued that a possible phase transition at temperature around $T\sim 0.2-0.3$ where heat conduction is normal above this temperature while anomalous below this temperature exists~\cite{Gendelman2000prl}. Although this phase transition statement was challenged as a finite size effect~\cite{Yang2005prl}, a later work supported this phase transition conjecture after deriving a similar temperature dependent thermal conductivity as $\kappa(T)\propto e^{-T/A}$~\cite{Pereira2006prl}.

Most recently, the 1D coupled rotator lattice re-attracts much attention due to the new finding of simultaneously existing normal diffusion of momentum as well as heat energy~\cite{Li2014arxiv}. It is argued that the normal behavior might be due to the reduced number of conserved quantities which is unique for periodic interaction potentials~\cite{Das2014arxiv,Spohn2014arxiv}. But the new debate is that whether the stretch is conserved or not in 1D coupled rotator lattice. The investigation of 1D coupled rotator lattice will be the key to unravel the true mechanism behind the connection between momentum conservation and normal or anomalous heat conduction. Therefore, it is the right time to revisit the temperature dependence of thermal conductivity of 1D coupled rotator lattice as a first step.

In this work, we will revisit the temperature dependence of thermal conductivities for the 1D coupled rotator lattice. We find that the temperature dependence is neither $\kappa(T)\propto e^{\Delta V/T}$ nor $\kappa(T)\propto e^{-T/A}$ as previously claimed~\cite{Giardina2000prl,Gendelman2000prl}. The actual temperature dependence is a power-law dependence as $\kappa(T)\propto T^{-3.2}$. The possible connection with the momentum diffusion of single kicked rotator or the Chirikov standard map has also been discussed. In order to determine whether there is a phase transition, we also present the temperature dependent thermal conductance at different system sizes. All the thermal conductances for different sizes collapse to the same value at low temperatures while approach to the power-law behavior as $\kappa(T)\propto T^{-3.3}$ at high temperatures. However, the crossover temperature decreases as the system size increases. This fact indicates that there is no phase transition between normal and anomalous heat conduction. In thermodynamical limit, the heat conduction is normal for all temperatures except the trivial zero temperature point. In Sec. II the lattice model will be introduced. Numerical results and discussions will be presented in Sec. III and the results will be summarized in Sec. IV.

\section{Model}
The Hamiltonian for the 1D coupled rotator lattice is defined as the following~\cite{Giardina2000prl,Gendelman2000prl}:
\begin{equation}\label{ham}
H=\sum^{N}_{i=1}\left[\frac{p^2_i}{2}+K[1-\cos{(q_{i+1}-q_i)}]\right]
\end{equation}
where $q_i$ and $p_i$ denote the displacement from equilibrium and momentum for $i$-th atom, respectively. The mass of the atom $m$ and the Boltzmann constant $k_B$ has been set into unity. The parameter $K$ with energy dimension represents the coupling strength of the inter-atom potential. Therefore, the system temperature $T$ can be rescaled by $T/K$ and we can also set $K=1$ for simplicity~\cite{Li2012rmp}. The equations of motion for $i$-th atom are
\begin{eqnarray}\label{equation-lattice}
\dot{q}_i&=&p_i \nonumber\\
\dot{p}_i&=&K\sin{(q_{i+1}-q_i)}-K\sin{(q_i-q_{i-1})}
\end{eqnarray}

At low temperature limit, the displacements are small values so that $|q_{i+1}-q_i|\ll 1$. The Hamiltonian can be expanded into the Taylor series as:
\begin{eqnarray}\label{rotator-low-T}
H&=&\sum^{N}_{i=1}\left[\frac{p^2_i}{2}+\frac{(q_{i+1}-q_i)^2}{2}\right.\nonumber\\
&-&\left.\frac{(q_{i+1}-q_i)^4}{4!}+\frac{(q_{i+1}-q_i)^6}{6!}-...\right]
\end{eqnarray}
This Hamiltonian will approach to the integrable Harmonic lattice only at zero temperature $T=0$. The first stable nonlinear potential term will be the sextic potential.

On the other hand, at infinite high temperature limit $T=\infty$, the rotator lattice will approach to another integrable system consisting of $N$ independent and isolated free particles as
\begin{equation}
H=\sum^{N}_{i=1}\frac{p^2_i}{2}
\end{equation}
This is because the kinetic energy is proportional to the temperature as $\left<p^2_i\right>=T$ and the potential energy in Eq. (\ref{ham}) is confined by the cosine function.

Without getting into numerics, we can get a qualitative picture of the thermal conductivity of 1D coupled rotator lattice as the function of temperature. The thermal conductivity $\kappa(T)$ will diverge in the low temperature limit approaching to the harmonic limit and will decay to zero in the high temperature limit as approaching to the unconnected $N$ free particles. As a general picture, the thermal conductivity $\kappa(T)$ will decrease as the temperature increases.

\begin{figure}
\includegraphics[width=\columnwidth]{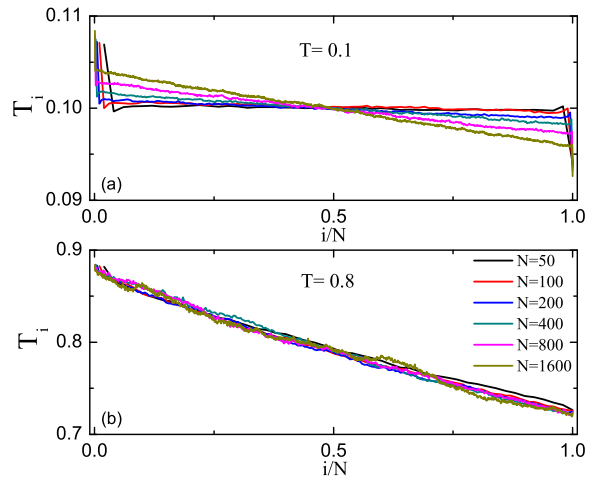}
\vspace{-0.5cm} \caption{\label{fig:T-profile}
(color online). Temperature profiles for the 1D coupled rotator lattice at (a) $T=0.1$ and (b) $T=0.8$. The lattice sizes are $N=50,100,200,400,800$ and $1600$ and the color rule of the lines are the same for both (a) and (b). The first and last atom are put into contact with a Langevin heat bath with temperature set as $T_{L/R}=T(1\pm\Delta)$ where $\Delta=0.1$ here.}
\end{figure}

In non-equilibrium numerical simulations, we put the first and last atom into contact with Langevin heat bath. The temperatures for left and right heat bath are set as $T_{L/R}=T(1\pm \Delta)$ where $T$ denotes the average temperature and $\Delta = 0.1$ gives rise to the temperature gradient along the lattice. The fixed boundary conditions with $q_0=q_{N+1}=0$ are also been applied.

\section{Results and Discussions}

Before we discuss the results of thermal conductivities, we first show the temperature profiles. In Fig. \ref{fig:T-profile}, the temperature profiles at two different temperatures $T=0.1$ and $T=0.8$ are plotted. The lattice sizes are chosen as $N=50,100,200,400,800$ and $1600$. For relative low temperature at $T=0.1$, the temperature jumps at two boundaries are obvious. However, the temperature jumps are reduced with the increase of lattice size $N$ as can be seen in Fig. \ref{fig:T-profile}(a). In Fig. \ref{fig:T-profile}(b) where the temperature is relatively high at $T=0.8$, all the temperature profiles collapse to the same straight line as the temperature jumps are very small for all lattice sizes.

In order to obtain the temperature dependence of thermal conductivities, we need first define the way how $\kappa(T)$ can be calculated numerically. We notice that the temperature profiles all collapse to the same straight line if the temperature jumps can be ignored at high temperatures or long lattice sizes. It is appropriate to define the thermal conductivity $\kappa(T)$ as:
\begin{equation}\label{kappa-def}
\kappa(T)=\frac{JN}{T_L-T_R}
\end{equation}
where $J=\left<J_i\right>$ is the average heat flux along the lattice in the stationary state and the local heat flux $J_i$ is defined as $J_i=-p_i K \sin{(q_{i+1}-q_i)}$ via the energy continuity equation.

\begin{figure}
\includegraphics[width=\columnwidth]{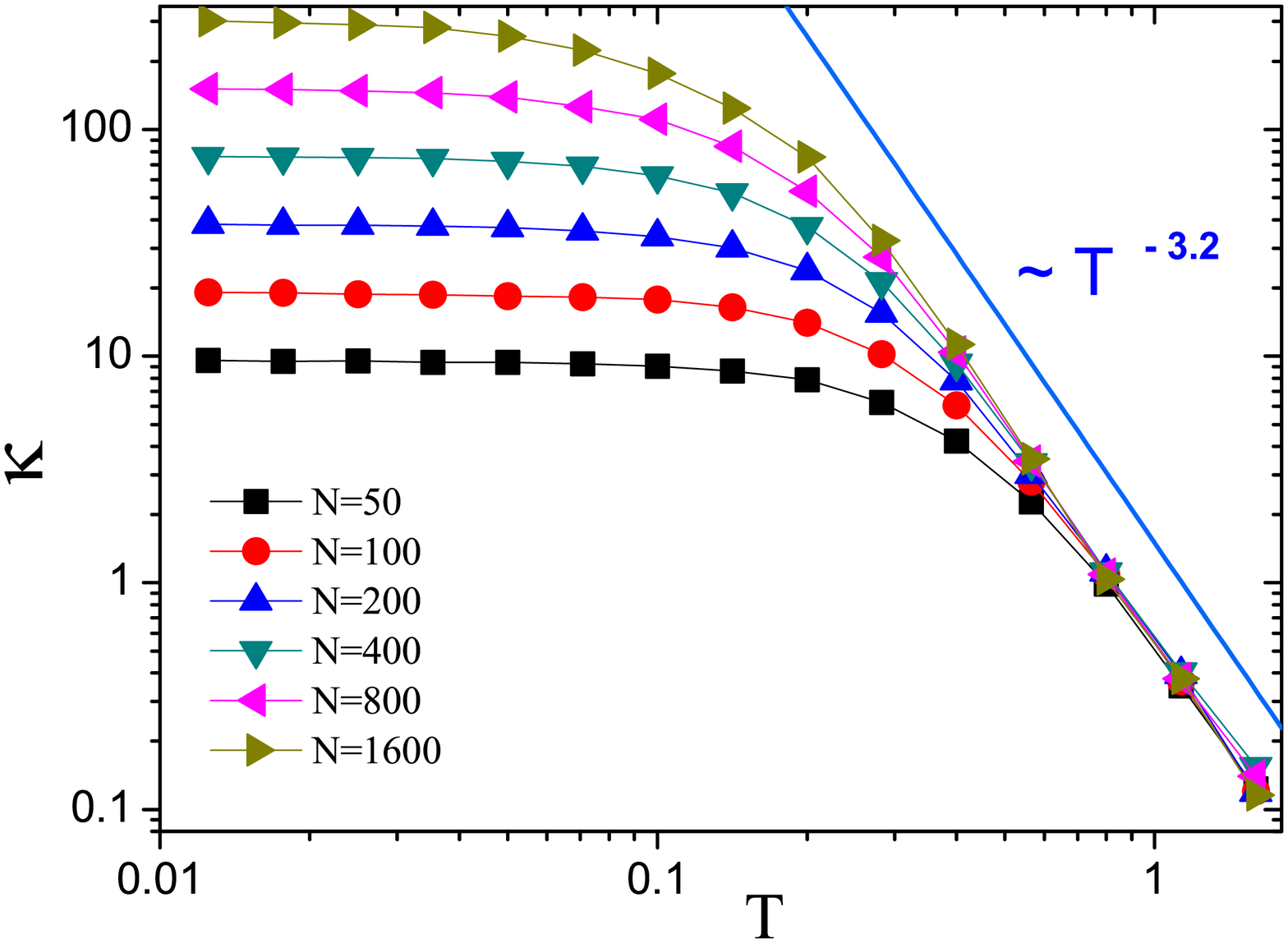}
\vspace{-0.5cm} \caption{\label{fig:kappa}
(color online). Thermal conductivities $\kappa(T)$ as the function of temperature for different lattice sizes $N=50,100,200,400,800$ and $1600$. The straight line is proportional to $T^{-3.2}$ which describes the temperature behavior of $\kappa(T)$ at high temperature region very well.}
\end{figure}

In Fig. \ref{fig:kappa}, the thermal conductivities $\kappa(T)$ as the function of temperature are plotted for different lattice sizes $N=50,100,200,400,800$ and $1600$. At high temperature region, all the thermal conductivities $\kappa(T)$ for different lattice sizes collapse together indicating the saturation of thermal conductivities as the increase of lattice size. This is characteristic for lattices with normal heat conduction. As the temperature decreases, the thermal conductivities first increases and then becomes flat. This is because the phonon mean free paths are getting longer as the temperature reduces and the ballistic regime will be approached if the phonon mean free paths are longer than the lattice size. The boundary jumps will dominate the temperature profiles as in Fig. \ref{fig:T-profile}(a) and the definition of $\kappa(T)$ of Eq. (\ref{kappa-def}) will no longer reflect the actual thermal conductivities.

\begin{figure}
\includegraphics[width=\columnwidth]{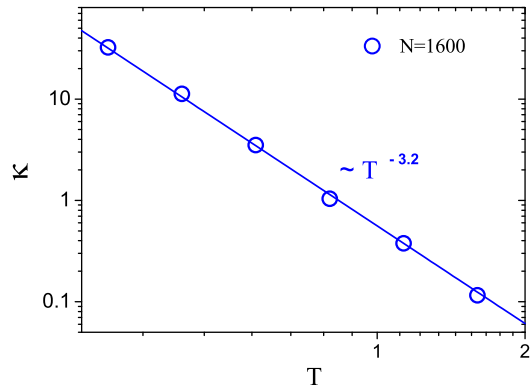}
\vspace{-0.5cm} \caption{\label{fig:kappa-T_1600}
(color online). Thermal conductivities $\kappa(T)$ as the function of temperature for lattice size $N=1600$. The data are taken from Fig. \ref{fig:kappa} and can be best fitted to be a power-law dependence as $\kappa(T)\propto T^{-3.2}$.}
\end{figure}

At high temperatures, it is clearly seen that the thermal conductivities follows a power-law dependence as $\kappa(T)\propto T^{-3.2}$. For the longest size we considered here as $N=1600$, this power-law behavior can be best fitted for more than two orders of magnitudes for the $\kappa(T)$ value as shown in Fig. \ref{fig:kappa-T_1600}. This also explains the poor fitting of the $\kappa(T)\propto e^{\Delta V/T}$ dependence in Ref. \cite{Giardina2000prl} and the $\kappa(T)\propto e^{-T/A}$ dependence in Ref. \cite{Gendelman2000prl}.

This power-law dependence of $\kappa(T)\propto T^{-3.2}$ cannot be explained by the effective phonon theory which is able to predict the temperature dependent thermal conductivities for other typical 1D nonlinear lattices such as FPU-$\beta$ lattice and generalized nonlinear Klein-Gordon lattices~\cite{Li2006epl,Li2007epl,Li2007pre,Li2009jpsj,Li2010prl,Li2012aip,Li2013pre,Yang2014pre}. According to the effective phonon theory, the thermal conductivity for low temperature rotator lattice with Hamiltonian of Eq. (\ref{rotator-low-T}) can be derived as
\begin{equation}\label{kappa-ept}
\kappa(T)\propto \frac{1}{\epsilon} \propto T^{-2}
\end{equation}
where $\epsilon$ is the nonlinearity strength with the following temperature dependence
\begin{equation}
\epsilon\propto \frac{\left<(q_{i+1}-q_i)^6\right>}{\left<(q_{i+1}-q_i)^2\right>}\propto T^2
\end{equation}
at low temperature region. Here we consider the sextic potential term as the lowest nonlinear term because the dynamics governed by the negative quartic potential term is unstable. Therefore, the actual temperature behavior of $\kappa(T)\propto T^{-3.2}$ is steeper than the prediction of $\kappa(T)\propto T^{-2}$ from effective phonon theory.

Although it is difficult to unravel the exact physical mechanism behind the power-law dependence of thermal conductivities for coupled rotator lattice, it is very helpful to look into the transport properties of the single kicked rotator (the Chirikov standard map)~\cite{Chirikov1979pr}. As the name indicates, the coupled rotator lattice is a kind of connected kicked rotators. The equations of motion for the single kicked rotator are
\begin{eqnarray}\label{equation-single}
p_{n+1}-p_{n}&=& K \sin{(q_n)}\nonumber\\
q_{n+1}-q_{n}&=& p_{n+1}
\end{eqnarray}
where $q_n$ and $p_n$ denote the coordinate and momentum after $n$-th kick.

\begin{figure}
\includegraphics[width=\columnwidth]{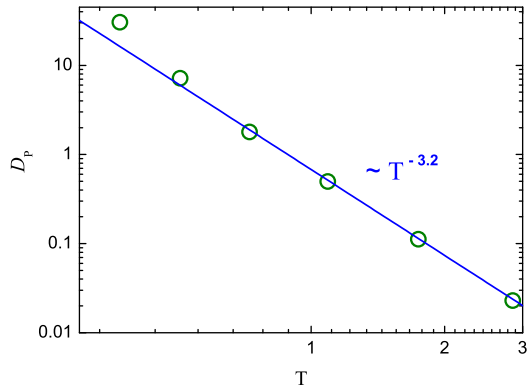}
\vspace{-0.5cm} \caption{\label{fig:Dp}
(color online). Momentum diffusion constant $D_P$ as the function of temperature for 1D coupled rotator lattice. The numerical data are obtained via equilibrium MD simulations as in Ref. \cite{Li2014arxiv}. The solid line of $T^{-3.2}$ is guided for the eye.}
\end{figure}

The variation of momentum $p$ is unbounded and can be characterized by a normal diffusion~\cite{Chirikov1979pr,MacKay1984pd}
\begin{equation}
\left<\Delta p^2(t)\right>\sim D t
\end{equation}
where $D$ is the diffusion constant. This is similar to the normal momentum diffusion for 1D coupled rotator lattice.
According to Ref. \cite{Chirikov1979pr,MacKay1984pd}, the diffusion constant $D$ depends on the coupling strength $K$ as
\begin{eqnarray}\label{dif-single}
D &\propto& (K-1.2)^3,\,\,\, 1.2<K<4 \\
D &\propto& K^2,\,\,\,K>4
\end{eqnarray}
where $1.2$ is the chaos threshold for the kicked rotator.

If one assume the transport properties of coupled rotator lattice are the same as that of single kicked rotator, one would expect that the energy diffusion constant $D_E$ as well as the momentum diffusion constant $D_P$ of coupled rotator lattice should also follow the same dependence as that of single kicked rotator as in Eq. (\ref{dif-single}). To translate the $K$ dependence into $T$ dependence, we notice that the parameter $K$ in single kicked rotator of Eq. (\ref{equation-single}) plays the same role as that in the coupled rotator lattice as in Eq. (\ref{equation-lattice}). As we have discussed above, the dynamics of coupled rotator lattice can be rescaled with $T/K$. The $K$ dependence should be inversely proportional to the $T$ dependence. Low $K$ value region corresponds to the high $T$ region and vice verse.  And for normal heat conduction, the thermal conductivity $\kappa$ is proportional to the energy diffusion constant $D_E$. This will finally give rise to the prediction of temperature dependent thermal conductivity $\kappa(T)$ of coupled rotator lattice from the analogy with single kicked rotator:
\begin{eqnarray}
\kappa(T) &\propto& T^{-2},\,\,\,\mbox{low} \,\,T\\
\kappa(T) &\propto& T^{-3},\,\,\,\mbox{high} \,\,T
\end{eqnarray}

As a consistence check, we plot the momentum diffusion constant $D_P$ for the coupled rotator lattice as the function of temperature in Fig. \ref{fig:Dp}. The same temperature dependence of $D_P\propto T^{-3.2}$ has been obtained.

Therefore, the prediction of $\kappa(T) \propto T^{-3}$ with analogy to the single kicked rotator at high temperature or low $K$ region is close to the numerical observation of $\kappa(T) \propto T^{-3.2}$. This indicates there might be some deeper connection between the diffusion behavior of single kicked rotator and 1D coupled rotator lattice. In addition, the analogy analysis also predicts a $\kappa(T) \propto T^{-2}$ behavior at low temperature region which is the same as the prediction from the effective phonon theory of Eq. (\ref{kappa-ept}). The reason why this temperature behavior cannot be observed in numerical simulations might be due to the severe finite size effect at low temperature region as can be seen in Fig. \ref{fig:T-profile}(a).

\begin{figure}
\includegraphics[width=\columnwidth]{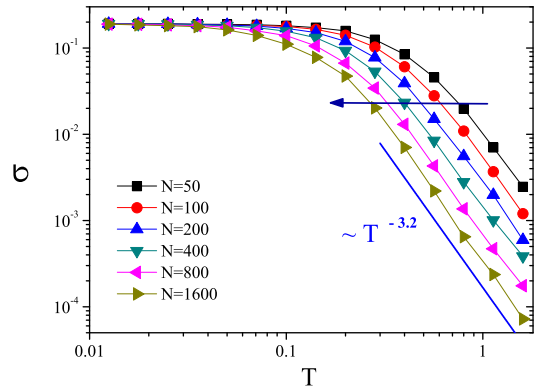}
\vspace{-0.5cm} \caption{\label{fig:conductance}
(color online). Heat conductance $\sigma$ as the function of temperature $T$ for different lattice size $N=50,100,200,400,800$ and $1600$. All the other parameters are the same as in Fig. \ref{fig:kappa}. The left-pointing arrow represents the trend of decreasing crossover temperature as the lattice size increases.}
\end{figure}

In the final part we will discuss the issue about the possible phase transition between normal and anomalous heat conduction for 1D coupled rotator lattice. As we have shown in Fig. \ref{fig:kappa-T_1600}, the actual temperature dependence of $\kappa(T)$ is a power-law dependence as $\kappa(T) \propto T^{-3.2}$. This indicates the analytic derivation of $\kappa(T)\propto e^{-T/A}$ in Ref. \cite{Pereira2006prl} originally for a lattice with pinned on-site potential can not be extended to 1D coupled rotator lattice.

On the other hand, the claim that the transition temperature is around $(0.2-0.3)$ in Ref. \cite{Gendelman2000prl} has been challenged in Ref. \cite{Yang2005prl} with numerical simulations of longer sizes. The possible phase transition could be a finite size effect. Here we reconfirm the finite size effect by giving a more illustrative picture in Fig. \ref{fig:conductance}. The heat conductance $\sigma\equiv\kappa(T)/N$ as the function of temperature has been plotted for different lattice sizes. At high temperature regions, all the heat conductances follow the same temperature behavior of $\sigma\propto T^{-3.2}$. As the temperature decreases, the phonon mean free path will overcome the lattice size and the heat conductance will saturate to a value determined by the harmonic limit of coupled rotator lattice. As can be seen from Fig. \ref{fig:conductance}, the heat conductance for short lattice will first bend and become flat as the temperature decreases. This crossover temperature decreases as the lattice size increases smoothly and no sign of phase transition can be observed. It is also noticed that the crossover temperature happens to be around $(0.2-0.3)$ for lattice sizes of a few thousands.

\section{Summary}
We have systematically investigated the temperature dependence of thermal conductivities of 1D coupled rotator lattice. The actual temperature dependence is a power-law dependence of $\kappa(T)\propto T^{-3.2}$ which is different with the observations of previous studies. The possible connection with the single kicked rotator or the Chirikov standard map has been discussed where a $\kappa(T)\propto T^{-3}$ dependence can be implied. Our results also reconfirm that the previously claimed possible phase transition should be a finite size effect.

\section{acknowledgments}
The numerical calculations were carried out at Shanghai Supercomputer Center. This work has
been supported by the NSF China with grant No. 11334007 (Y.L., N.L., B.L.), the NSF China with grant No. 11347216 (Y.L), Tongji University under Grant No. 2013KJ025 (Y.L), the NSF China with Grant No. 11205114 (N.L.), the Program for New Century Excellent Talents
of the Ministry of Education of China with Grant No. NCET-12-0409 (N.L.) and the
Shanghai Rising-Star Program with grant No. 13QA1403600 (N.L.).

\bibliographystyle{apsrev4-1}

\end{document}